\documentclass[ras]{agutex}

 \usepackage{graphicx}
 \usepackage{amsmath}
\setkeys{Gin}{draft=false}
\authorrunninghead{NEBEN ET AL.}

\titlerunninghead{MEASURING MWA BEAMPATTERNS WITH ORBCOMM SATELLITES}

\authoraddr{Corresponding author: A.R. Neben,
Kavli Institute for Astrophysics and Space Research, Massachusetts Institute of Technology, Cambridge, MA 02139, USA.
(abrahamn@mit.edu)}

\begin{document}

\title{Measuring Phased-Array Antenna Beampatterns with High Dynamic Range for the Murchison Widefield Array using 137 MHz ORBCOMM Satellites}

\author{A.~R.~Neben$^{1}$,
R.~F.~Bradley$^{2,3}$, 
J.~N.~Hewitt$^{1}$,
G.~Bernardi$^{8,11,18}$,
J.~D.~Bowman$^{4}$, 
F.~Briggs$^{5}$,
R.~J.~Cappallo$^{9}$, 
A.~A.~Deshpande$^{10}$, 
R.~Goeke$^{1}$,
L.~J.~Greenhill$^{8}$,
B.~J.~Hazelton$^{14}$, 
M.~Johnston-Hollitt$^{16}$,
D.~L.~Kaplan$^{15}$, 
C.~J.~Lonsdale$^{9}$, 
S.~R.~McWhirter$^{9}$,
D.~A.~Mitchell$^{6,18}$, 
M.~F.~Morales$^{14}$, 
E.~Morgan$^{1}$, 
D.~Oberoi$^{12}$, 
S.~M.~Ord$^{7,17}$,
T.~Prabu$^{10}$, 
N.~Udaya~Shankar$^{10}$, 
K.~S.~Srivani$^{10}$, 
R.~Subrahmanyan$^{10,17}$, 
S.~J.~Tingay$^{7,17}$, 
R.~B.~Wayth$^{7,17}$, 
R.~L.~Webster$^{13,17}$, 
A.~Williams$^{7}$, 
C.~L.~Williams$^{1}$}

$^{1}$Kavli Institute for Astrophysics and Space Research, Massachusetts Institute of Technology, Cambridge, MA 02139, USA\\
$^{2}$Dept. of Electrical and Computer Engineering, University of Virginia, Charlottesville, VA, 22904\\
$^{3}$National Radio Astronomy Obs., Charlottesville, VA\\
$^{4}$School of Earth and Space Exploration, Arizona State University, Tempe, AZ 85287, USA\\
$^{5}$Research School of Astronomy and Astrophysics, Australian National University, Canberra, ACT 2611, Australia\\
$^{6}$CSIRO Astronomy and Space Science (CASS), PO Box 76, Epping, NSW 1710, Australia\\
$^{7}$International Centre for Radio Astronomy Research, Curtin University, Bentley, WA 6102, Australia\\
$^{8}$Harvard-Smithsonian Center for Astrophysics, Cambridge, MA 02138, USA\\
$^{9}$MIT Haystack Observatory, Westford, MA 01886, USA\\
$^{10}$Raman Research Institute, Bangalore 560080, India\\
$^{11}$Square Kilometre Array South Africa (SKA SA), Cape Town 7405, South Africa\\
$^{12}$National Centre for Radio Astrophysics, Tata Institute for Fundamental Research, Pune 411007, India\\
$^{13}$School of Physics, The University of Melbourne, Parkville, VIC 3010, Australia\\
$^{14}$Department of Physics, University of Washington, Seattle, WA 98195, USA\\
$^{15}$Department of Physics, University of Wisconsin--Milwaukee, Milwaukee, WI 53201, USA\\
$^{16}$School of Chemical \& Physical Sciences, Victoria University of Wellington, Wellington 6140, New Zealand\\
$^{17}$ARC Centre of Excellence for All-sky Astrophysics (CAASTRO)\\
$^{18}$Department of Physics and Electronics, Rhodes University, PO Box 94, Grahamstown, 6140, South Africa

\begin{abstract}
Detection of the fluctuations in 21 cm line emission from neutral hydrogen during the Epoch of Reionization in thousand hour integrations poses stringent requirements on calibration and image quality, both of which necessitate accurate primary beam models. The Murchison Widefield Array (MWA) uses phased array antenna elements which maximize collecting area at the cost of complexity. To quantify their performance, we have developed a novel beam measurement system using the 137 MHz ORBCOMM satellite constellation and a reference dipole antenna. Using power ratio measurements, we measure the {\it in situ} beampattern of the MWA antenna tile relative to that of the reference antenna, canceling the variation of satellite flux or polarization with time. We employ angular averaging to mitigate multipath effects (ground scattering), and assess environmental systematics with a null experiment in which the MWA tile is replaced with a second reference dipole. We achieve beam measurements over 30 dB dynamic range in beam sensitivity over a large field of view (65\% of the visible sky), far wider and deeper than drift scans through astronomical sources allow. We verify an analytic model of the MWA tile at this frequency within a few percent statistical scatter within the full width at half maximum. Towards the edges of the main lobe and in the sidelobes, we measure tens of percent systematic deviations. We compare these errors with those expected from known beamforming errors. 

\end{abstract}

\begin{article}

\section{Introduction}

The prospects of studying the formation of the first structures in the universe at $z\sim6$ and earlier with 21 cm hydrogen emission have driven investment in a new generation of low frequency radio astronomy instruments (see \citet{FurlanettoReview, miguelreview, PritchardLoebReview, aviBook, zaroubi} for reviews). The extreme surface brightness sensitivity required to detect the 21 cm signal in the presence of galactic, extragalactic, and thermal noise backgrounds has pushed this generation of experiments into an untested regime. Uncertain primary beams and source catalogs, in addition to wide fields of view, complicate analysis and demand new methods of calibration, imaging, and primary beam characterization. 

A number of experiments now operating, such as the Murchison Widefield Array (MWA) \citep{lonsdale09,tingay13,mwascience}, the Donald C. Backer Precision Array for Probing the Epoch of Reionization (PAPER) \citep{paperinstrument, parsons14}, and the LOw Frequency Array (LOFAR) \citep{lofar}, as well as demonstrator instruments like the MIT Epoch of Reionization Array (MITEoR) \citep{zheng14}, next-generation experiments like the Hydrogen Epoch of Reionization Array (HERA; \citet{PoberNextGen}; http://reionization.org), and future instruments like the Square Kilometer Array \citep{ska} have opted for large arrays (hundreds of elements) of non-pointing or only coarsely-pointing antennas, attempting to balance collecting area and cost considerations. For all these experiments, \textit{in situ} high-fidelity primary beam characterization remains a major challenge given the high dynamic range \citep[e.g.][]{AaronSensitivity, beardsley13, nithya13, PoberNextGen} thought necessary to reveal the cosmological 21 cm signal. 

Recovering the 21 cm signal in the presence of strong foregrounds is made easier by taking advantage of an effective containment of smooth spectrum foregrounds in a Fourier space region knows as the ``wedge'', despite the frequency-dependent response of the interferometer \citep{Dattapowerspec,X13, PoberWedge,MoralesPSShapes, VedanthamWedge, nithya13, CathWedge, AdrianWedge1, AdrianWedge2}. However, frequency dependent systematics due to insufficiently accurate primary beam modeling for calibration or primary beam correction may cause foreground leakage out of this compact region. This would shrink the region within which a cosmological power spectrum measurement can be made and thus lower the significance of a detection \citep{PoberNextGen}. In fact, \citet{nithya2015} show that most pernicious for such measurements is sky emission from large zenith angles, even near the horizon, just where beampatterns are most difficult to model. Ultimately, the full polarization response may be needed to best model and subtract polarized sources that can leak a sinusoidal frequency signal into Stokes I due to their Faraday rotation \citep{jelic2010,moore2013}. Recent measurements \citep{giannisurvey, moore2013,moore2015,asad2015} indicate that most of the high rotation measure sources are, however, largely depolarized at low frequency.

Though brought to the forefront again by 21 cm science, primary beam measurements have a long history in radio astronomy and electrical engineering. Radio astronomers typically rely upon celestial radio sources with known flux densities to measure beampatterns as the sources trace out cuts through the beam \citep[e.g.,][]{nithyaVLA}. If the beam is narrow enough for the sky to appear as a single point source, knowledge of its flux density is not needed to measure relative beam sensitivity along its track \citep[e.g.][]{colegate14}, though combining tracks from different sources, or using fields with multiple sources requires accurate knowledge of their relative fluxes. Indeed, the wide fields of view of dipole elements and uncertainties in low frequency source catalogs make this analysis difficult and entangled with calibration \citep{jacobs2013}. Further, the lack of axial symmetry in non-dish antennas around the antenna pointing direction makes a complete beampattern impossible to measure from just a handful of cuts. Relying only on the weaker assumption of $180^\circ$ rotation symmetry, \citet{pober12} present an interferometric beam measurement technique making use of celestial sources with unknown flux densities, assuming the data are already calibrated.  

In this paper we pursue an alternate approach based on probe signals from satellites. Satellite-based antenna beampattern measurements have many clear advantages over astronomical sources. Satellites are substantially brighter, and thus dominate in otherwise crowded fields and probe deep into beam sidelobes. They also make many cuts through the beam over the course of many orbits, due to precession. \citet{brueckmann63} exploited these advantages in early satellite measurements of antenna beampatterns. Lingering issues such as varying antenna pointing and plane of polarization (due to Faraday Rotation or simple projection effects) and time-varying satellite transmitter power may be solved through use of a simple, well-understood reference antenna and power ratio measurements, which measure the relative beampattern of the Antenna-Under-Test (AUT) \citep[eg.][]{fukao85, law97, hurtado2001, vlbamemo}. This approach has been used to great effect in holographic antenna measurements \citep{rochblatt92, harp2011, lasenby85, godwin86, deguchi93}, with at most a handful of satellites.

As a first step towards antenna beam measurements exploiting all these advantages for 21 cm cosmology and for the MWA, we develop a prototype of a novel beam measurement system using probe signals from the 137 MHz ORBCOMM satellite constellation and a reference dipole antenna. The precession of these low Earth orbit satellites and their sheer number ($\sim30$) yield $65\%$ coverage of the visible sky (limited by satellite coverage at our Green Bank site) at $2^\circ$ resolution in a single day. Early tests and demonstrations of our initial concept were presented by \citet{ries07, BradleyAndRies2008,CzekalaAndBradley2010,CzekalaAndBradley2010_2,aasposter}, and recently,  \citet{zheng14} have demonstrated a simple implementation of this concept in a working interferometer. We use an ORBCOMM interface box to determine satellite transmission frequencies on the fly, allowing us to take advantage of even more of the ORBCOMM signals in an automated manner.

In this work, we present the full working power pattern mapping system, as well as an error analysis of environmental systematics such as multipath (ground scattering). We expect the lessons learned about these systematics to inform future beam measurements and array calibration methods employing probe signals carried by remote-controlled drones or satellites. 

We use our working beam measurement system to make the first precision measurements of an MWA tile in a deployment-style environment over a large fraction of sky. Each MWA tile is a $4\times4$ grid of bowtie dipoles, optimized to have a broad frequency response over 80--300 MHz, whose signals are combined in a delay-line beamformer. This design results in a large collecting area per tile and a beam narrow enough to steer away from the bright galactic disk and terrestrial RFI, but adds model complexity and uncertainty near the edges of the main beam and in the sidelobes (grating lobes). Indeed the MWA tile design poses simulation challenges due both to potential dipole cross coupling effects as well as the large number of degrees of freedom which must by simulated (hundreds of frequencies, dozens of pointings, and fine angular resolution). Note, though, that current processing of MWA data utilizes simulated beampatterns, and experiments such as that presented in this work aim more to assess their validity than to replace them with measurements.

Early beampattern measurements of deployed prototype MWA tiles using source drift scans \citep{bowman07} and anechoic chambers \citep{williamsthesis2012} revealed rough agreement with models. Yet $\sim1$ dB deviations were observed throughout the main lobe and $3-5$ dB deviations in sidelobes, highlighting the need both for better modeling and better control of measurement systematics. Later, \citet{sutinjo2015} found that upwards of 200 MHz, interactions between antennas necessitate more complex modeling than simple multiplication of a Hertzian dipole beampattern by the array factor. We are interested here in characterizing the deviations from ideality at lower frequencies where the simple model is more likely to hold. At some level, the beams are expected to be corrupted by beamforming amplitude and phase errors, finite ground screen effects, as well as any dipole-dipole interactions. We compare our measurements with an empirical budget of beamforming errors due to dipole phase and gain mismatching which we present in a separate paper (Neben et al., in prep).

In Section \ref{sec:measurementsystem} we discuss our beam measurement system including the ORBCOMM satellites, our reference antenna, and our data acquisition system. We discuss our data analysis pipeline in Section \ref{sec:analysis} and also present results of a null experiment in which a second reference antenna is used as the Antenna-Under-Test. We present beam measurements of our MWA tile and compare with models in Section \ref{sec:mwatile}, and conclude with discussion and conclusions in Section \ref{sec:discussion}.

\section{Measurement System}
\label{sec:measurementsystem}

\subsection{Overview}

\begin{figure}
\includegraphics[width=7in]{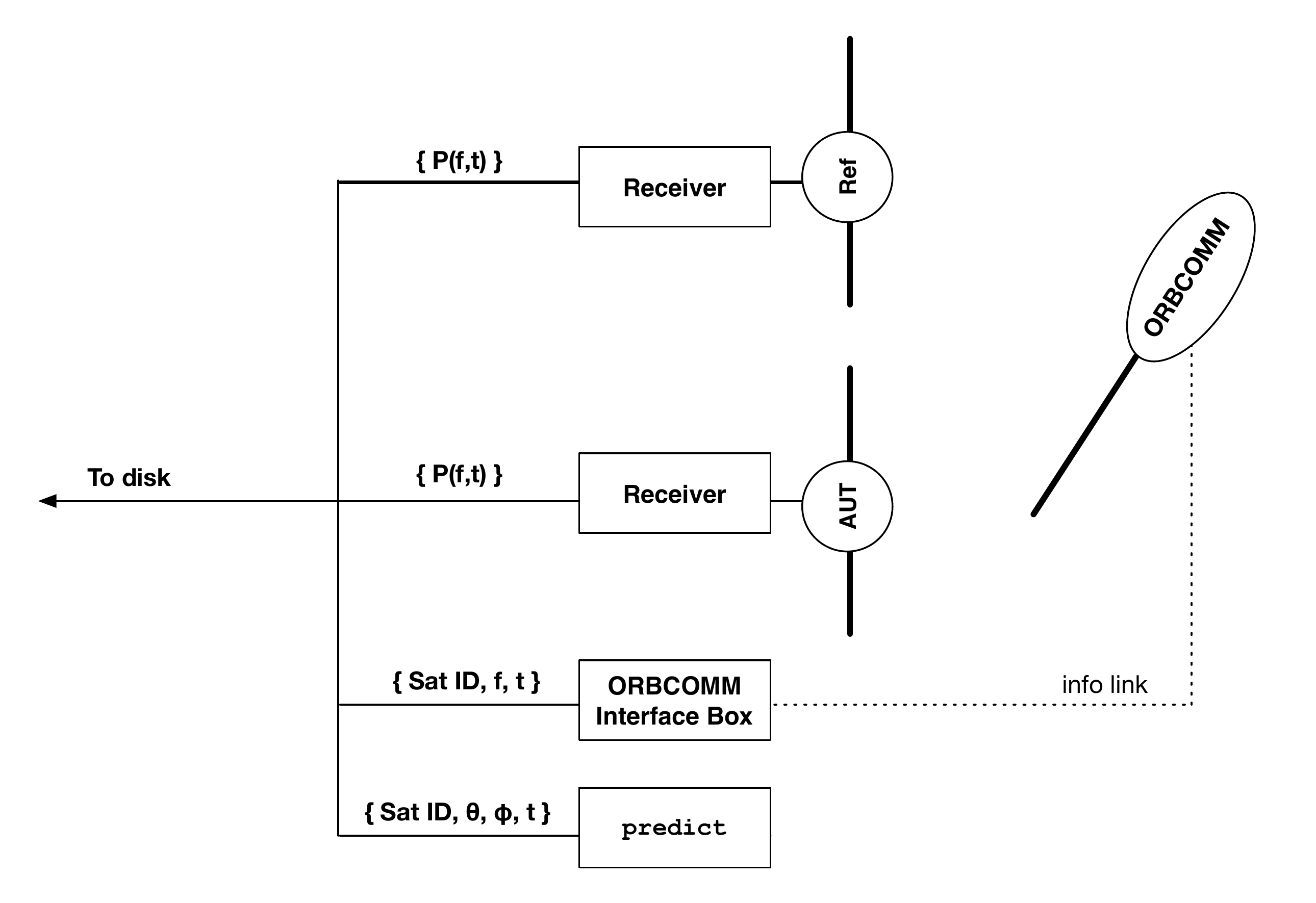}
\caption{Simplified diagram of our beam measurement system and data flow. ORBCOMM satellite signals are received by our Reference Antenna and the AUT, each passing through the receiver chain described in Sec. \ref{sec:measurementsystem} which outputs a power spectrum with 2 kHz resolution between 137--138 MHz every $\sim$200 ms. At each time step, a copy of {\tt predict} running on our data acquisition computer outputs the positions and IDs of all ORBCOMM satellites above the horizon, while transmission frequencies are logged by our ORBCOMM Interface box. }
\label{fig:systemoverviewdiagram}
\end{figure}

Figure \ref{fig:systemoverviewdiagram} shows a schematic diagram of our antenna setup and measurement chain. A reference antenna (see Section \ref{sec:refant}) and an Antenna-Under-Test (AUT) both pick up transmitted signals from a passing ORBCOMM satellite, which are mixed down, sampled, and fast fourier transformed by our data acquisition system (see Section \ref{sec:daq}), and finally the power as a function of frequency from both antennas is saved to disk at every time step. Note that in practice, both our antennas are dual-polarization, necessitating two additional measurement chains like these. Additionally, an ORBCOMM Interface Box (typically supplied to commercial users of the ORBCOMM constellation) interfaces with each passing satellite and outputs its identifier (which allows precise prediction of the satellite's location using orbital data) and frequency channel (which specifies which $\sim$20 kHz wide ORBCOMM band between 137--138 MHz the satellite is transmitting on). 

This work was performed at the National Radio Astronomy Observatory site in Green Bank, West Virginia, located in the US National Radio Quiet Zone. Though the ORBCOMM satellites dominate the radio sky in the 137--138 MHz band, wide band interference from terrestrial radio transmitters causes saturation problems in other geographic locations. Our AUT and reference antenna are positioned 50m apart on a North-South baseline located at (38.429348$^\circ$N, -79.845737$^\circ$E), and aligned to an accuracy of about a degree, as confirmed by Google Earth (http://www.google.com/earth) imagery. The Green Bank site is not perfect, however, and the surrounding hills and radio telescopes raise concerns of  shadowing and multipath effects. We assess these with a null experiment in which the beampattern of a known reference-style dipole is used as the AUT (see Section \ref{sec:nullexpt}).

In this work we measure the angular response of each instrumental polarization (NS and EW) to unpolarized radiation, leaving for future work measurement of the full polarized beampatterns. While it is not obvious that fully polarized satellite probe signals suffice to measure unpolarized beampatterns, we show in Appendix \ref{sec:measurementappendix} that this is in fact possible if both our reference antenna and our AUT have the same polarization response. 

\subsection{ORBCOMM Satellite Constellation}
\label{sec:orbcomms}

ORBCOMM Inc. operates a constellation of $\sim30$ communications satellites in low Earth orbit (altitude $\sim800$km) designed for users requiring low baud rate communication with remote sites. The satellites provide excellent Earth-coverage and near continuous transmission, predominantly occupying orbital planes with inclinations between $\pm45^\circ$ and 10 narrow ($\sim20$ kHz wide) subbands in the 137--138 MHz band. An advantage of these satellites over higher altitude satellites (such as the GPS constellation) for beampattern measurements is that good sky coverage is achieved far more quickly due to the shorter orbital period and rapid orbital precession resulting from the lower altitude. In particular, sky coverage at our Green Bank site is limited only by absence of satellites with inclinations greater than $45^\circ$. The information content of the transmitted satellite signals is irrelevant for our purpose and is lost in the RMS power measurements of our data acquisition system (see Section \ref{sec:daq}). 

Each satellite's frequency is relatively stable over days, but shifts periodically to avoid interference within the constellation. There are typically several ORBCOMM satellites above the horizon at any given time, and while we can easily compute their positions using published orbital elements, we must determine which frequencies correspond to which satellites. \citet{zheng14} use interferometric phases to identify and exclude times when more than one satellite is present. We are able to take advantage of \textit{all} satellite passes using an ORBCOMM User Interface Box (typically provided by ORBCOMM Inc. to users) connected to a separate antenna, whose debug port logs the satellite number and frequency band occupied by each passing satellite. 

During data collection, we record the satellite positions using the Linux program {\tt predict} \footnote{http://www.qsl.net/kd2bd/predict.html} which numerically integrates the orbits using orbital elements (TLE) data published by Celestrak\footnote{http://www.celestrak.com/NORAD/elements/orbcomm.txt} daily. We run two copies of {\tt predict} in live multi-satellite tracking mode on our data acquisition (DAQ) computer, and query them for the angular positions of all satellites currently above the horizon whenever a satellite power measurement is made (typically every 200ms). We save this information with the recorded satellite power data. See Section \ref{sec:daq} for a detailed description of our data acquisition system.

\subsection{Reference Antenna}
\label{sec:refant}
\begin{figure}
\includegraphics[height=1.5in]{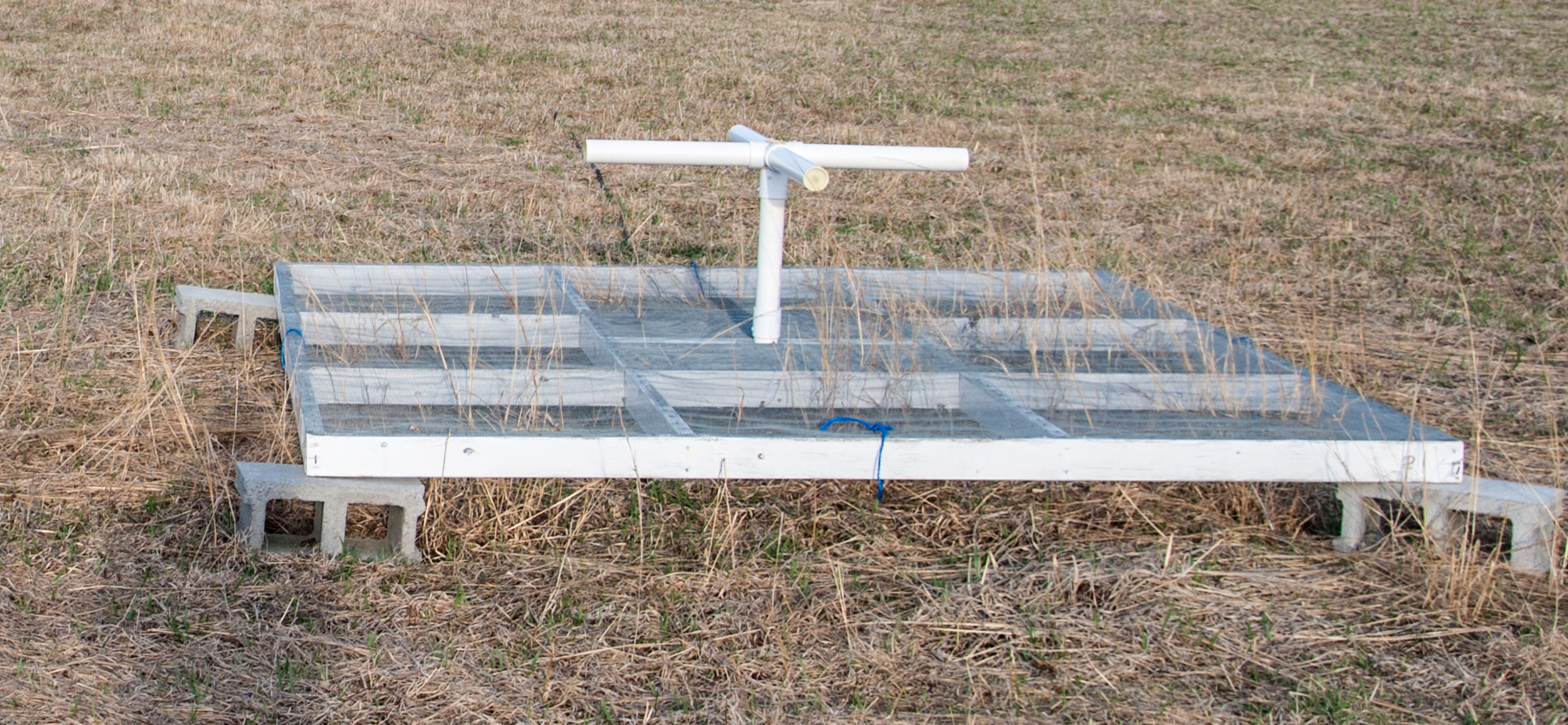}
\includegraphics[height=1.5in]{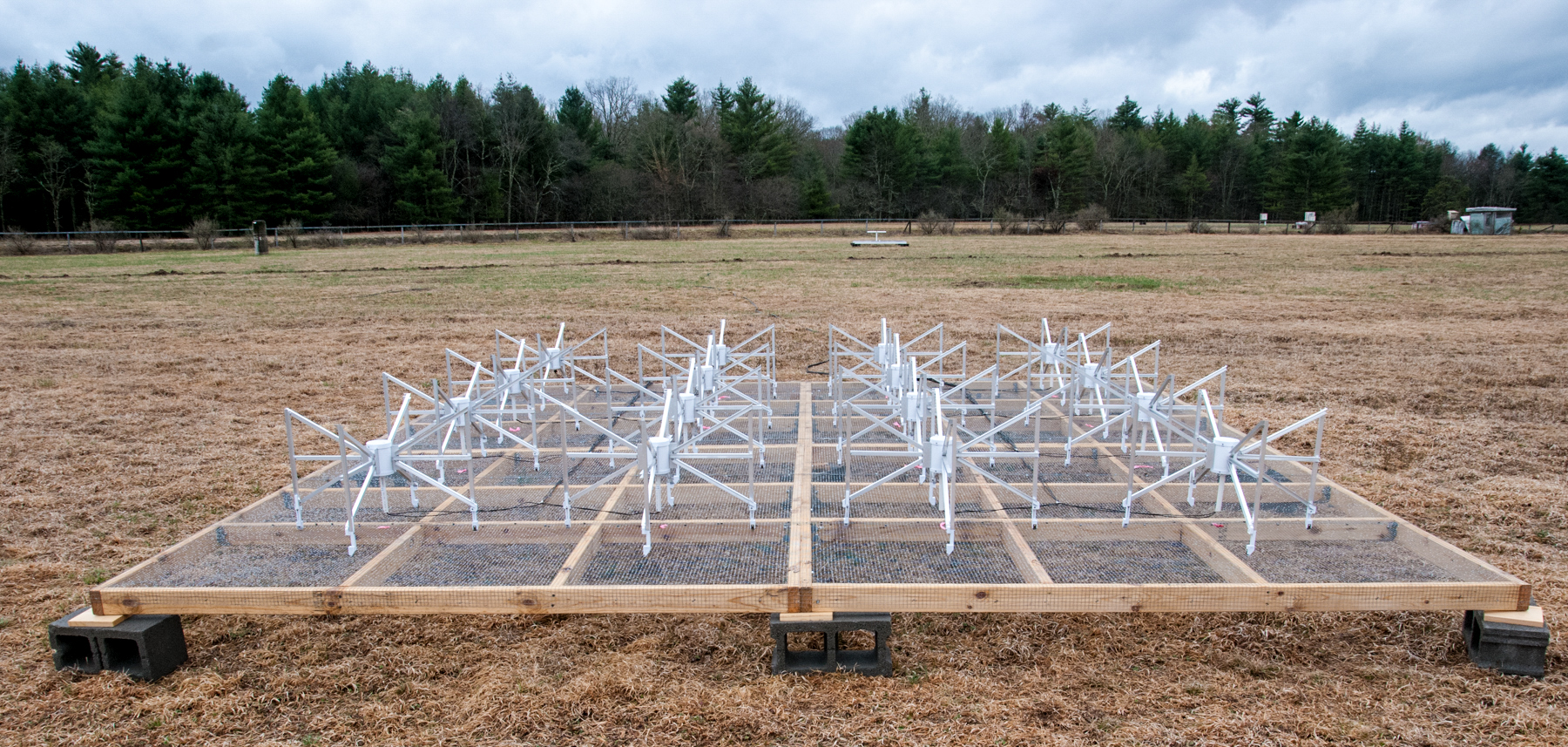}
\caption{Reference antenna on 2 m $\times$ 2 m ground screen (left) (see Section \ref{sec:refant}) and MWA antenna tile on 5 m $\times$ 5 m ground screen (right) (see Section \ref{sec:mwatile}) deployed at the National Radio Astronomy Observatory--Green Bank.}
\label{fig:antennas} 
\end{figure}

Our reference antenna is a simple dual-polarization dipole mounted above a 2 m $\times$ 2 m ground plane, elevated 48.3 cm above the soil (see left panel of Figure \ref{fig:antennas}). The antenna itself is made from 90.4 cm long and 0.5 inch diameter copper tubing, and is encapsulated in 2 in diameter Schedule 40 PVC tubing. The beampattern of the dipole was derived from an electromagnetic analysis of the physical structure over a finite ground plane using CST Microwave Studio \footnote{https://www.cst.com/Products/CSTMWS}. The soil is modeled as a cube of lossy dielectric 3.05 m on a side having a relative permittivity of 13 and electrical conductivity of 0.005 S/m. 

\subsection{Data Acquisition Hardware}
\label{sec:daq}

Our receiver produces a mean power measurement in each frequency channel in each of two instrument polarizations (NS and EW) for each of our two antennas (the AUT and the reference antenna) every $\sim$200 ms. The raw signals are first mixed with a 127 MHz local oscillator down to an intermediate frequency of $\sim$10.5 MHz, then simultaneously sampled at 2 MHz, digitally mixing them down to the 0--1 MHz baseband (i.e. one of the Nyquist zones falls in the frequency range 0--1 MHz).  A 12-bit ADC acquires a burst of 51,200 samples on which 50 FFTs of length 1024 samples are performed, effectively covering the 137--138 MHz bandwidth with a resolution 2 kHz (cf. $\sim$20 kHz bandwidths of ORBCOMMS). From these measured voltage Fourier amplitudes $\widetilde{V}_{\mathrm{ant},\mathrm{pol},i}(f)$, where $i$ runs from 1 to 50, the RMS powers $\langle|\widetilde{V}_{\mathrm{ant},\mathrm{pol},i}(f)|^2\rangle_i$ are saved to disk for each antenna, polarization, and frequency channel along with a list of all satellites currently above the horizon and their angular positions (obtained from {\tt predict}, see Section \ref{sec:orbcomms}). A full complex polarization analysis of the AUT beam would be possible given measurements of voltage cross powers $\langle \widetilde{V}_{\mathrm{ant},\mathrm{pol},i}(f)\widetilde{V}_{\mathrm{ant'},\mathrm{pol'},i}^*(f)\rangle_i$, though this places stringent requirements on instrumental phase stability and we thus reserve it for future work.

\section{Data Analysis}
\label{sec:analysis}
\subsection{Processing Satellite Passes}
\label{sec:processingsatellitepasses}
The first step is to determine each satellite's transmission frequency during each pass. We have developed a script to extract this information from the captured debug output of the ORBCOMM User Interface Box, which periodically ``syncs'' with passing satellites and logs their identifiers and transmission frequencies. We assume a time window of 30 minutes over which the recorded frequencies are valid, and use this mapping of satellites and time windows in the next step of the analysis.

Next, we manipulate the satellite power data. At each $\sim$200 ms time step, a beam measurement can be made at the position of each satellite above the horizon (each of which is transmitting on a different, but now known frequency). This is done separately for EW and NS instrumental polarizations. The measured AUT and reference dipole powers in each satellite's frequency band are determined by integrating the measured RMS band powers over the central 15 kHz of the satellite's signal band. The background power level is estimated as the minimum of the received power level at start and end of each pass (when the satellite is below the horizon), and data within 20 dB of that floor are rejected. This ensures that the bias on measured beampattern due to the sky noise is less than 1\%. 

As a heuristic description of our measurements assuming the satellite signals are unpolarized, consider the received powers by the AUT and the reference antenna, $P_\mathrm{AUT}$ and $P_\mathrm{ref}$, in one instrumental polarization, and let $B_\mathrm{ref}$ and $B_\mathrm{AUT}$ be their unpolarized beam responses at the angular position of the satellite. Let $F$ be the incident flux from the satellite so that $P_\mathrm{ref}=B_\mathrm{ref}F$ and $P_\mathrm{AUT}=B_\mathrm{AUT}F$. Then it is clear than the AUT beam response in the satellite direction is given by
 \begin{equation}
\label{eqn:beammeasurement}
B_\mathrm{AUT}=P_\mathrm{AUT}B_\mathrm{ref}/P_\mathrm{ref}
\end{equation}
This is essentially our analysis, done separately for EW and NS oriented antennas. We show in Appendix \ref{sec:measurementappendix} that the polarization of satellite signals does not affect Equation \ref{eqn:beammeasurement}, assuming both antennas have the same polarization response.

\subsection{Forming Power Patterns}
\label{sec:formingpowerpatterns}

To form power patterns, we grid the measured AUT beam values from many satellite passes into equal solid angle cells on the sky using the HEALPix software package \citep{healpix} to (1) facilitate comparison with model power patterns; (2) average over short-timescale fluctuations due to noise and multipath effects; and (3) facilitate rejection of outliers in each sky pixel due to rare saturation issues. We choose a cell size of 1.8 deg (nside=32) which results in $\sim$5 satellite passes and $\sim75$ measurements in each pixel per day, out of which $\sim5$ typically fall outside of the central 90\% and are rejected as outliers. This results in few percent precision on our measured beams and sufficient resolution to resolve features of interest except within several degrees of the the MWA beam nulls, where the beam changes on sub-pixel scales.  Normalizing the measured beam might be done by rescaling it to peak at unity, though we opt for a less noisy normalization by fitting for a rescale factor to best match the measured beam to the normalized analytic model within $\sim10$ degrees of boresight. 

\subsection{Assessing Systematics with a Null Experiment}
\label{sec:nullexpt}

We characterize systematics using a null experiment, in which we use a second reference-style antenna as the AUT. The beampatterns of the two reference antennas will deviate from each other due to environmental effects (e.g., multipath effects or shadowing) and instrumental non-idealities (e.g., alignment errors or imperfections in the ground screen, soil, or dipole itself). Our null experiment will effectively measure the ratio of these two antenna beampatterns, and thus, the level at which they deviate from each other. We interpret this measure as a rough proxy for deviation of each antenna away from the ideal electromagnetic model.

\begin{figure}
\includegraphics[width=6.5in]{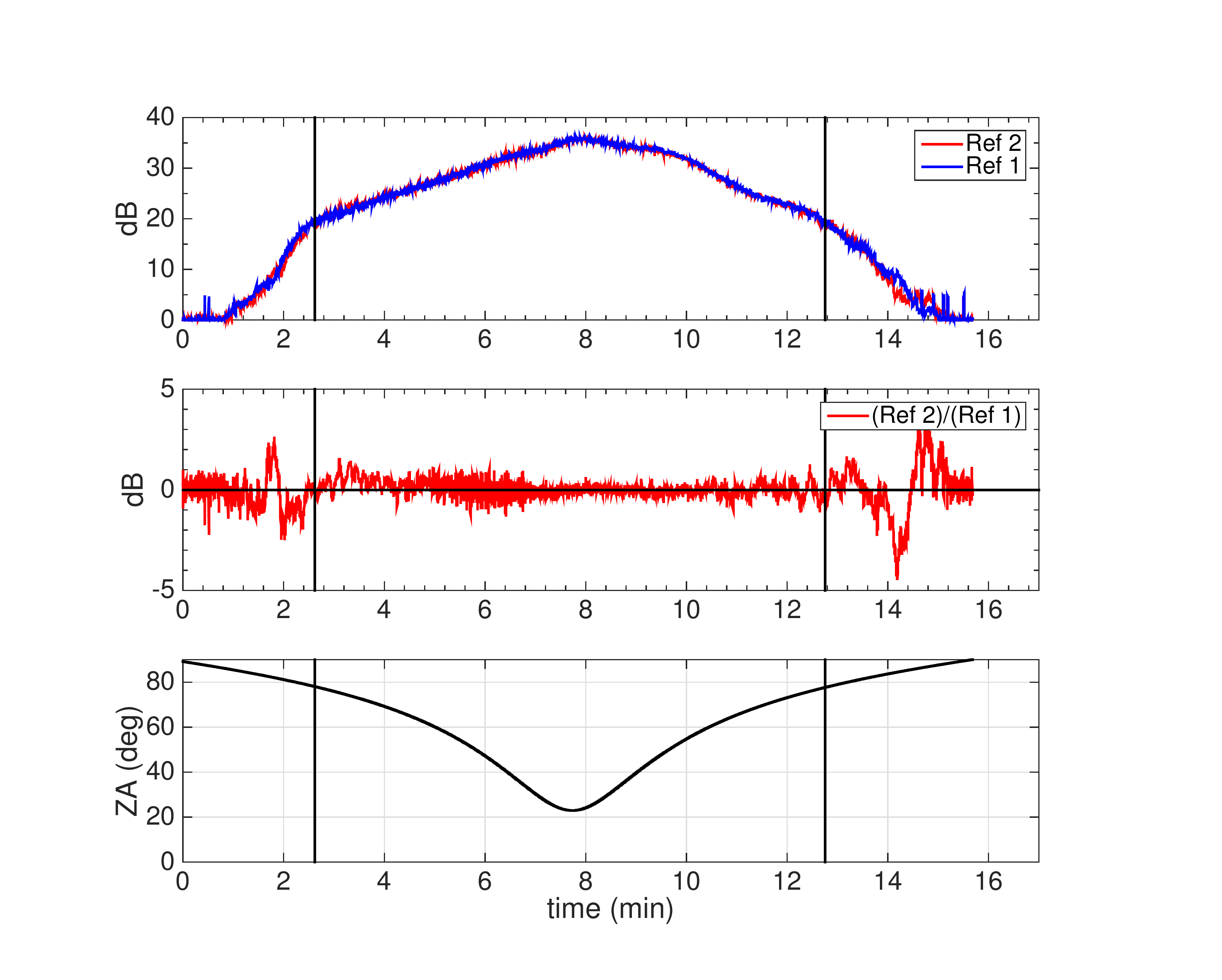}
\caption{Analysis of a typical satellite pass from our null experiment, in which the AUT is replaced by a reference-style antenna. The satellite rises out of the galactic background power (see Fig. \ref{fig:skynoise}), passes high in the sky, then drops below the horizon over the course of roughly 15 minutes (top panel). Both curves are in units of dB relative to the background level estimated at the beginning of the pass. Outside the region enclosed by vertical lines, the signal to background ratio is smaller than 20 dB, meaning that the satellite signal received by each antenna is corrupted at the few percent and larger level by diffuse galactic power.  The fluctuations in the ratio of the two antenna responses (middle panel) are typically at the $\pm0.5$dB level  and are due to multipath reflections and polarization mismatch, not receiver noise. We also plot the satellite zenith angle (bottom).}
\label{fig:satellitepass}
\end{figure}

Figure \ref{fig:satellitepass} shows a satellite pass from this setup in detail. Over the course of 15 minutes, the satellite rises out of the background, passes through the visible sky, and falls below again (shown in the top panel). The units are dB relative to the background level (shown below to be predominately diffuse galactic emission), and we mark with vertical lines the region within which the received power is more than 20 dB stronger than the background. If the AUT beampattern is different from that of the reference dipole, beam measurements outside this region will suffer systematics at a level of a few percent and larger due to the diffuse galactic emission received in addition to that from the satellite. We opt to simply avoid this region in lieu of subtracting a background model. The ratio of the two reference antenna powers (shown in the middle panel) is mostly consistent with unity up to $0.5-1$ dB statistical scatter, systematic biases of comparable magnitudes are apparent at large zenith angles. As this is just one satellite pass, it is difficult to draw general conclusions about which regions of the sky are least or most susceptible to such biases. Only after gridding many satellite passes together does a fuller picture emerge. Note, though, that given the brightness of the ORBCOMM satellites and the averaging discussed in \ref{sec:formingpowerpatterns}, we interpret the statistical fluctuations as a combination of multipath reflections (different at the two antenna locations) and polarization mismatch, not as receiver noise. 

\begin{figure}
\includegraphics[width=5in]{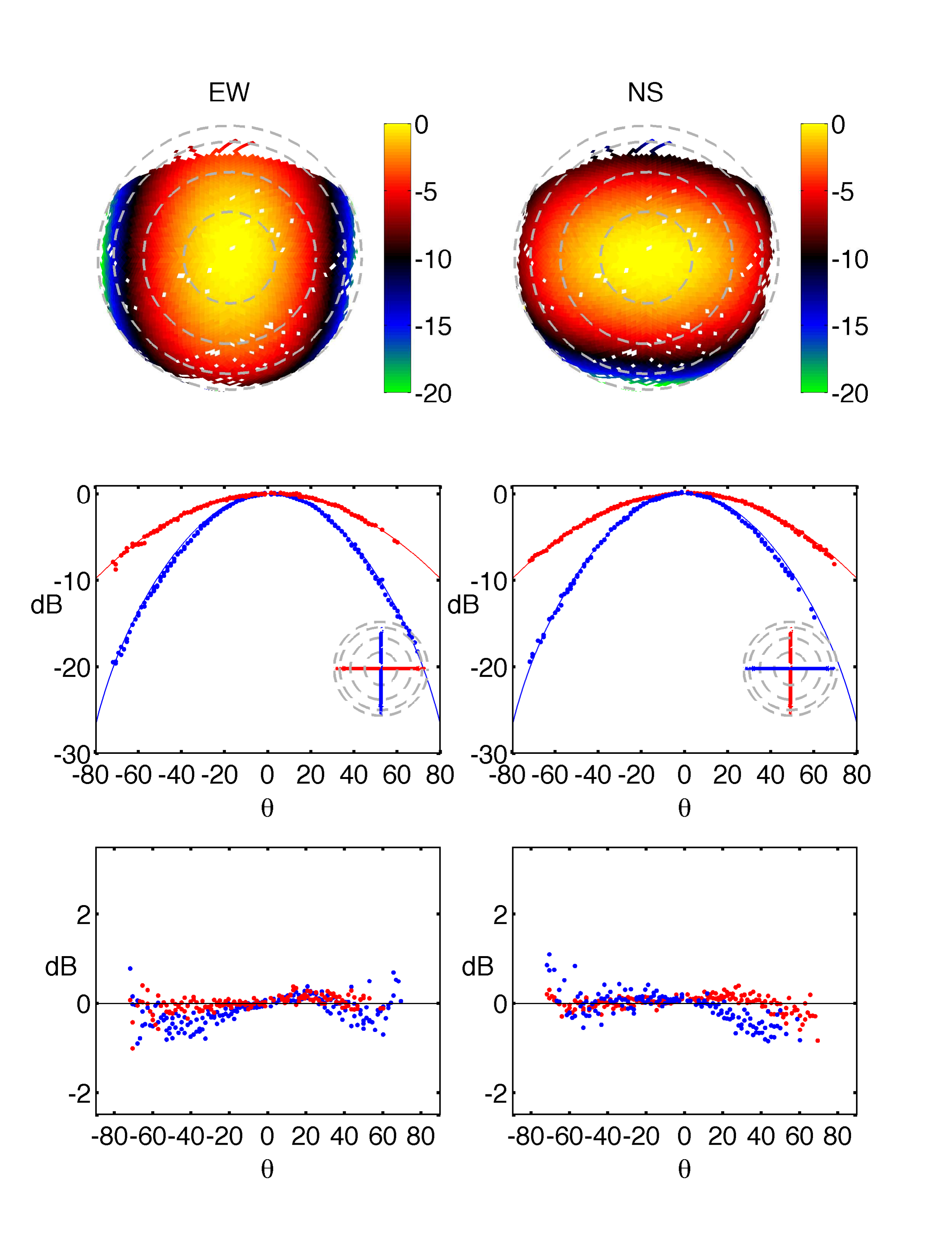}
\caption{Results from our null experiment in which the AUT is another reference antenna. Beams of the EW (NS) oriented dipoles are shown in the left (right) column. The measured AUT beampattern is plotted in dB relative to its boresight gain (top). These maps are in sine projection with North at the top and East at the right. Dashed circles mark 20, 40, 60, and 80 degrees from zenith. We also show measured and model beampatterns (middle) and deviations from the model (bottom) on slices through the E (red) and H (blue) antenna planes.}
\label{fig:null2map}
\end{figure}

To characterize the behavior of these fluctuations as they manifest in power patterns, we combine 296 satellite passes in the null experiment configuration recorded over 32 hours (spread over 4 days) into a measured beampattern of the reference antenna (Figure \ref{fig:null2map}). The measured reference antenna beampattern is consistent with our numerical model within few percent statistical scatter within 20 degrees of zenith, and shows modest systematic trends at the 10\% level farther out suggestive of a few degree rotational misalignment. This level of agreement sets an upper limit on beam measurement systematics due to environmental effects and instrumental non-idealities as discussed above. We thus interpret these results as measurements of the accuracy and precision of our beam measurement system in its current configuration. In Sec. \ref{sec:discussion}, we discuss approaches to mitigating these systematics further.

\begin{figure}
\includegraphics[width=5in]{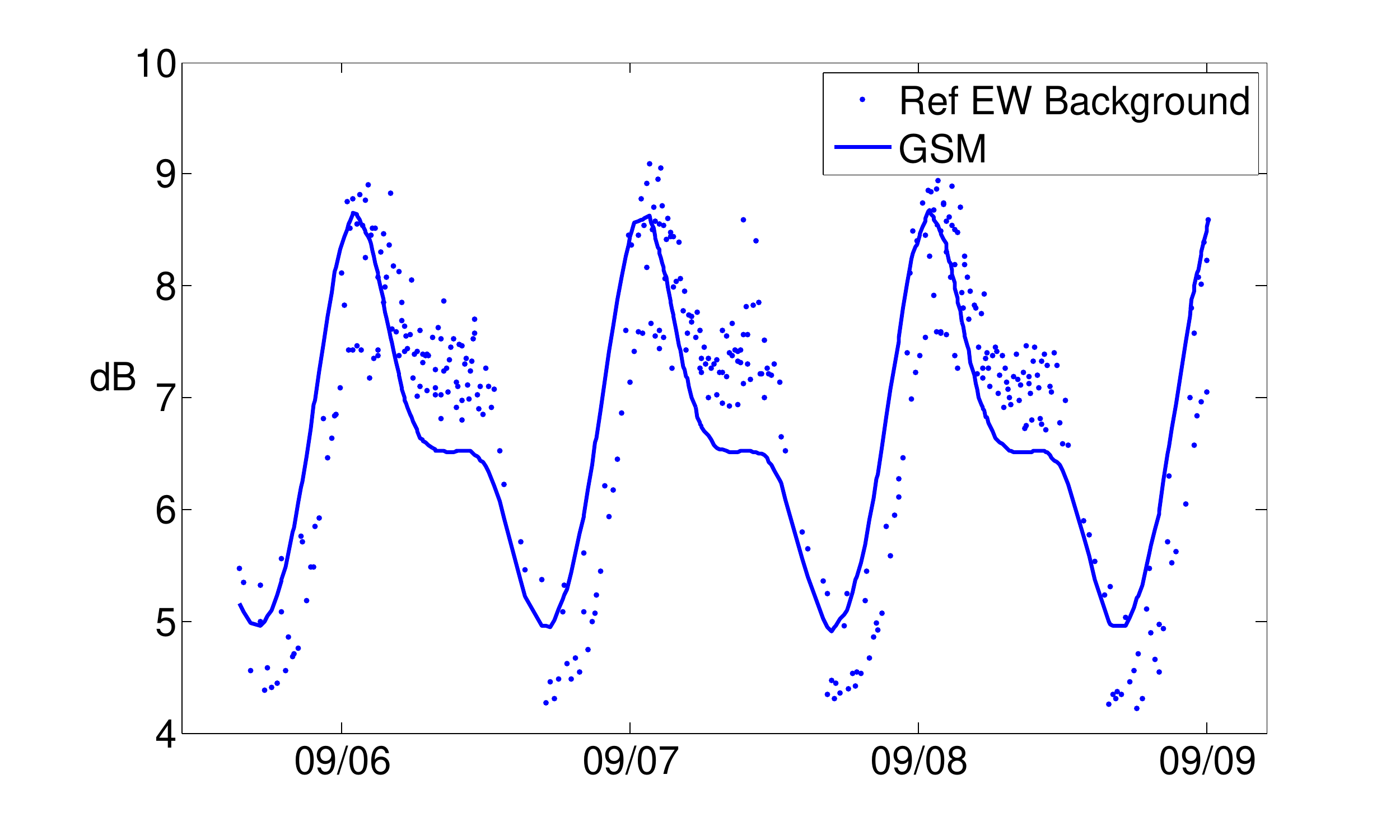}
\caption{Observed reference antenna background levels measured at the beginning or end of each satellite pass, plotted with a model time-dependent background computed from the Global Sky Model (GSM) \citep{gsm} and the model reference antenna beampattern. The data agree with the GSM model given its stated $\pm10$\% accuracy.}
\label{fig:skynoise}
\end{figure}

As a check on the reliability of our background estimation (used only to identify and avoid times of significant background power relative to satellite power), we plot in Figure \ref{fig:skynoise} the observed background level as a function of time versus that predicted from the Global Sky Model (de Oliveira Costa 2008) and our model reference antenna beampattern. Even neglecting the sun, the observed background estimates agree qualitatively with the GSM at the $\pm0.5$ dB level, close to its stated accuracy of $\pm10\%$. Note that slight phase and amplitude disagreements at this level are expected as the GSM is generated through interpolation between sky maps at other frequencies, and errors are thus correlated over large scales. That the observed background levels are  roughly consistent with the predicted galaxy power demonstrates that the ORBCOMM satellites are spaced sufficiently far apart in their orbits and frequency bands so as to not overlap in time, which is crucial for our experiment.

\section{MWA Antenna Tile}
\label{sec:mwatile}

Having quantified the accuracy and precision of our setup, we proceed to a study of the beampattern of an MWA antenna tile (hereafter MWA tile). The MWA tile consists of a 4$\times$4 grid of dual-polarization bowtie dipoles whose signals are combined in a delay-line beamformer. The dipoles are vertical bowties with dimensions 74 cm across and 84 cm on the diagonal, optimized to have a broad frequency response in the 100--200 MHz band. They are mounted on a 5 m $\times$ 5 m ground screen attached to a leveled wooden frame approximately 20 cm above the ground. The center-to-center dipole spacing is  1.1m and the center-to-ground-screen distance is 30 cm. For this experiment, we construct the ground screen out of 5 pieces of wire mesh (19 gauge, 0.5'' spacing) which are crimped together every 5 inches to form a constant potential surface connected to earth ground. Each dipole has a 20 dB LNA with integrated balun. The beamformer sums 16 dipole signals for each of the two polarizations (NS and EW), producing a beam with Full-Width-at-Half-Maximum (FWHM) $\sim23^\circ$ at 137 MHz, with sidelobes at the $-20$ dB level. By digitally engaging delay lines in increments of $\sim$450 ps to each signal pathway, the beam may be steered far from zenith. The delays are engaged through simple digital control with no amplitude or phase calibration needed. Slight deviation from perfect gain and delay matching (see Sec. 4.3) across the beamformer pathways is one of the mismodeling effects this work will probe. Lastly, we note that beamformer adds 30 dB of gain to the summed signal, to which we add 36 dB of attenuation to avoid saturating the ADC in our receiver. 

\subsection{Model Beampatterns}
\label{sec:modelbeampatterns}

As 137 MHz is well below the half wavelength frequencies of the characteristic lengths of the MWA Dipole (202 MHz and 178 MHz), the Hertzian dipole model is expected to be valid, though deviations near the sidelobes would not be unexpected. The phased array and ground screen factors of the MWA tile are encapsulated in the array factor $\textbf{A}_{11}(\theta,\phi)=\textbf{A}_{22}(\theta,\phi+\pi/2)$ (see Appendix A),
\begin{equation}
\label{eqn:tilemodel}
|\textbf{A}_{11}(\theta,\phi)|^2 \propto \sin^2\left(kh\cos\theta\right)\left|\sum_{i=1}^{16}e^{i (\vec{k}\cdot \vec{x}_i+\eta_i)}\right|^2
\end{equation}
Here $h=0.3$ m is the height of the dipole midpoints off the ground screen, $\vec{k}$ is the direction of the satellite, $\eta_i$ is the phase delay applied by the beamformer, and $\vec{x_i}$ is the position of antenna $i$ on the grid in the $xy$ plane. The phase delay can be expressed as $\eta_i=2\pi fd_i\times$(435 ps) where $d_i$ is an integer between 0 and 31 which is specified when controlling the beamformer. Lastly, we use coordinates where $\hat{x}$ points towards the East, $\hat{y}$ points towards the North, and $\hat{z}$ points towards Zenith, and $\vec{k}=\frac{2\pi}{\lambda}(\sin\theta\sin\phi\hat{x}+\sin\theta\cos\phi\hat{y}+\cos\theta\hat{z})$. Combining Equations \ref{eqn:tilemodel}, \ref{eqn:rotmat}, and \ref{eqn:autbeam} gives our analytic MWA tile model. 

We also compare our measurements to the more advanced beam model presented by \citet{sutinjo2015} which includes dipole-dipole coupling effects and a numerical dipole model on realistically modeled soil. We use their average embedded element model, as work on a full electromagnetic coupling model is ongoing. Below $\sim180$ MHz, the corrections due to dipole coupling effects are small within the main lobe, but potentially observable in beam measurement extending into the sidelobes, like ours.

\subsection{Beampattern Measurements}

\begin{figure}
\includegraphics[width=5in]{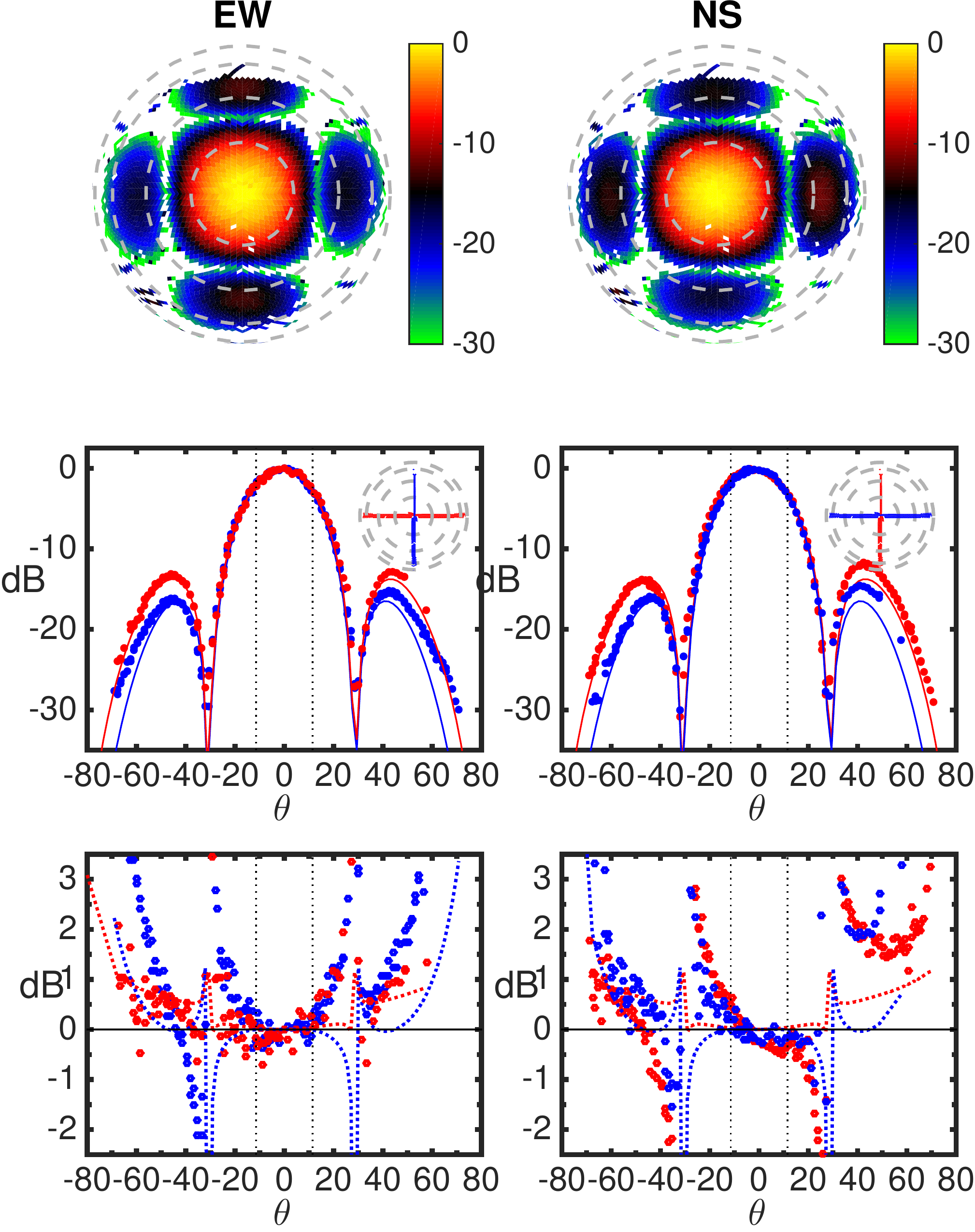}
\caption{The measured MWA tile beampattern for the zenith pointing is plotted in dB relative to its boresight gain (top). Beams of the EW (NS) oriented antennas are shown in the left (right) column. The maps are in sine projection with North at the top and East at the right. Dashed circles mark 20, 40, 60, and 80 degrees from zenith. We show measured and model beampatterns (middle) and deviations from the model (bottom) on slices through the E (red) and H (blue) antenna planes. Dashed lines in the bottom panel show the model of \citet{sutinjo2015} relative to the analytic model. Vertical dotted lines mark the model FWHM of $\sim$23 degrees at 137 MHz.}
\label{fig:zenithtilemap}
\end{figure}

Figure \ref{fig:zenithtilemap} shows our measured MWA tile beampattern (top panel) when pointed towards zenith, constructed from $\sim$400 satellite passes recorded over $\sim$4 days. The beam is plotted on the HEALPIX grid discussed in Sec. \ref{sec:formingpowerpatterns}. We also plot the measured beampattern and our analytic model on slices through these polar plots on the E and H (red and blue, respectively) antenna planes (middle panel), as well as the ratio of measured over analytic model beams (bottom panel). The numerical model of \citet{sutinjo2015} is plotted relative to the analytic model in the bottom panel for comparison (dashed lines), and should align with the data points if it explains the observed deviations. While measured beampatterns are often compared with models in simple beam sensitivity \textit{difference} plots, we view \textit{ratio} plots (i.e., differences of dB quantities) as more relevant given that primary beam sky weighting during both calibration and primary beam correction are multiplicative operations. Ratio plots also highlight the off-zenith regions where beams are typically most poorly modeled, the regions where foregrounds are most at risk of affecting EOR science, as discussed in Sec. 1.

Within the half power point ($\sim$12 degrees away from zenith) we observe agreement with our analytic model beam pattern within $\sim$5\% statistical fluctuations. Beyond that zenith angle, towards the edges of the main lobe and in the few percent sidelobes, we observe systematic deviations away from the model at the dB level. We discuss these deviations and their patterns further in Sec. \ref{sec:mwaerrors}. Both positive and negative trends are observed.

\begin{figure}
\includegraphics[width=5in]{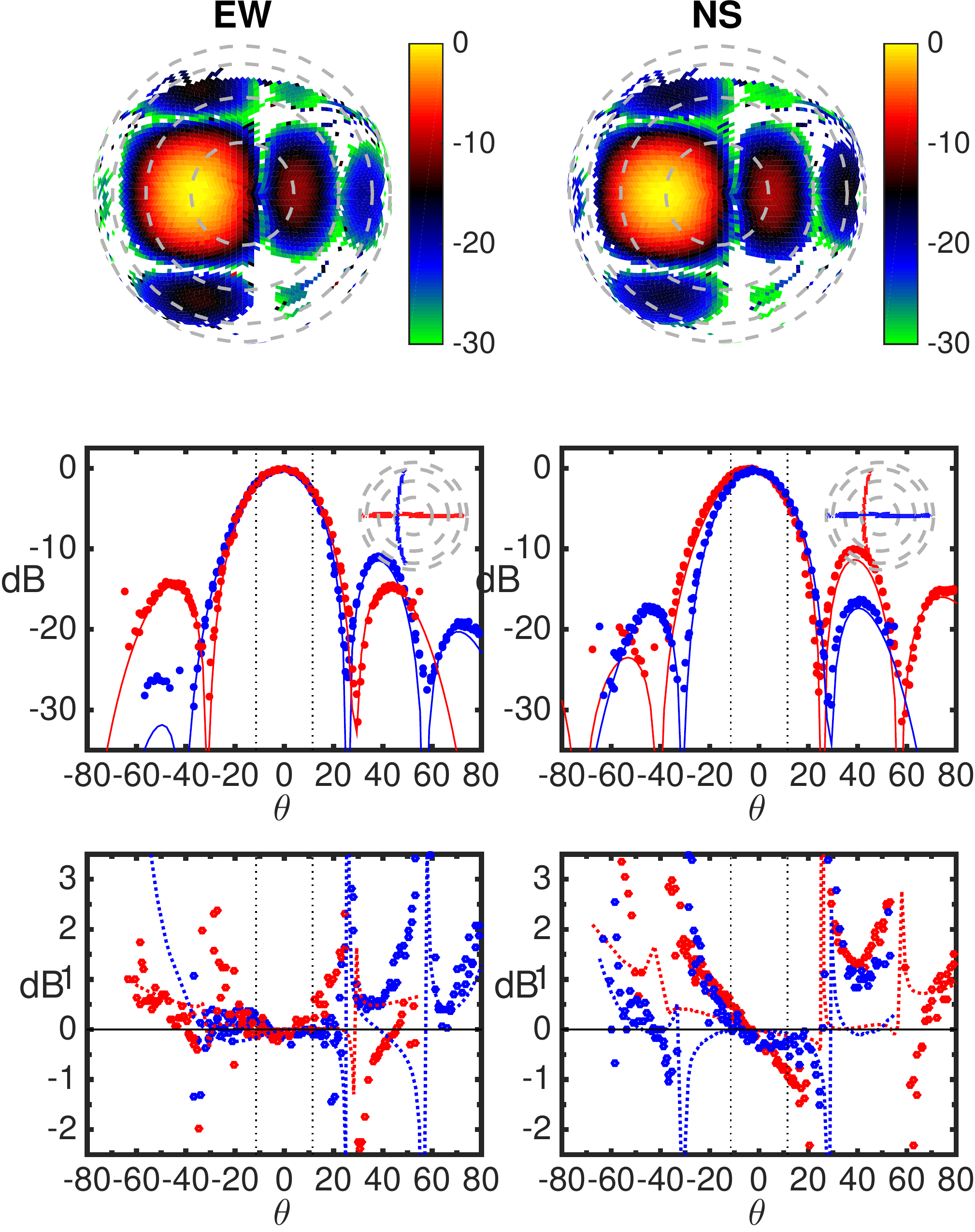}
\caption{Measured MWA tile power pattern for pointed 20 degrees West pointing. Same layout as Figure \ref{fig:zenithtilemap}, except here we plot beam slices through orthogonal planes through the main lobe (red and blue) which do not correspond to the E and H antenna planes because the direction of boresight is no longer in these planes.}
\label{fig:E03S00tilemap}
\end{figure}

\begin{figure}
\includegraphics[width=5in]{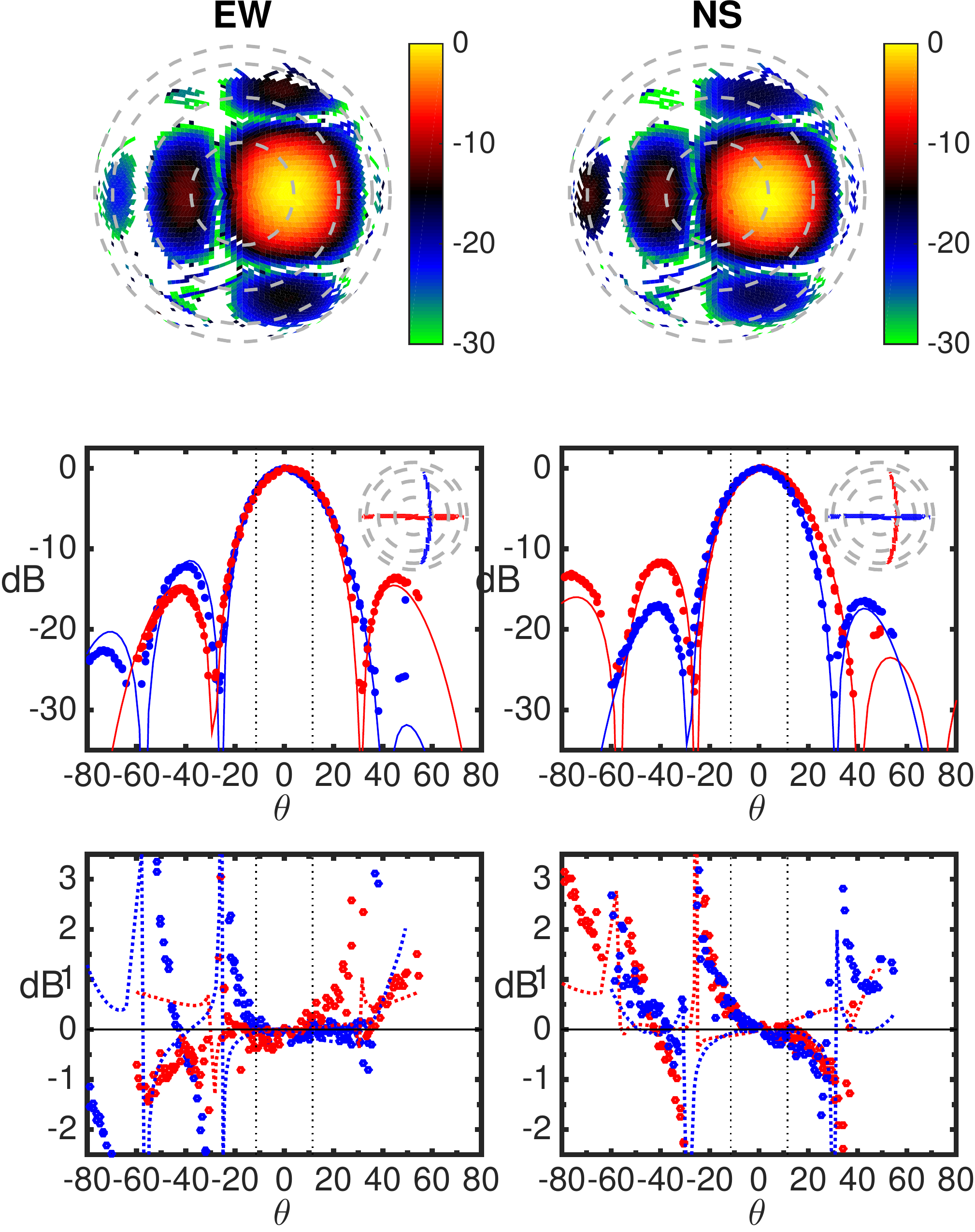}
\caption{Measured MWA tile power pattern for pointed 20 degrees East pointing. Same layout as Figure \ref{fig:E03S00tilemap}.}
\label{fig:W03S00tilemap}
\end{figure}

We have also measured the MWA tile power pattern at several off-zenith pointings, two of which are shown in Figures \ref{fig:E03S00tilemap} (20 degrees East) and \ref{fig:W03S00tilemap} (20 degrees West). For these pointings, the direction of boresight is not in the E and H antenna planes, so we instead rotate the antenna E and H planes to the boresight direction, yielding two orthogonal planes crossing the off-zenith main lobe. The level of deviation away from the model power patterns here is comparable with the Zenith pointing. 

\subsection{Error Analysis}
\label{sec:mwaerrors}

A detailed analysis of the causes and effects of beamforming errors in MWA tiles will be presented in a separate paper (Neben et al., in prep). In particular, that work is concerned with the finite precision of complex gain matching across the 16 beamformer signal paths (for each instrumental polarization) as well as tile rotation/tilt errors and dipole position errors. A budget of relevant systematics is established through laboratory measurements and compilation of manufacturer specifications and Monte Carlo simulations are run to propagate component uncertainties into direction-dependent beam power pattern uncertainties. Beam deviations at the level of 10--20\% are predicted in the sidelobes and near the edge of the main lobe, and significantly larger in the nulls. 

There are, however, several sources of error in the beam measurements presented in this work which are peculiar to our MWA tile at Green Bank. Our ground screen is elevated 20 cm off the ground on a wooden frame whereas the analytic and the cross-coupling models assume it is set on the ground. This is expected to affect the beam pattern at low elevations in particular, though in a symmetric manner. Non-coplanarity due to ground screen sag between its  frame supports also complicates the ground screen term in Eqn. \ref{eqn:tilemodel}, and may also introduce relative dipole tilts. Our ground screen is formed out of five rectangles of wire mesh, which are crimped together every 5". In contrast, the ground screen used in the deployed MWA tiles at the Murchison Radio Observatory in Western Australia are overlapped and welded, fixing a constant potential surface. 

Lastly slight tilts and rotations of our MWA tile due to our $\pm1.5^\circ$ alignment precision are larger than those affecting deployed MWA tiles. Such alignment errors are most prominent where the beam changes rapidly with angle as it does near the edges of the main lobe and in the sidelobes, and are difficult to correct for in our beam mapping experiment as they upset the polarization matching with the reference antenna. Numerical experiments suggest such alignment errors contribute systematics at the $\pm$20\% level.

Many of these errors will break the ideal symmetries of the tile beampatterns by introducing distortions and tilts of the main lobe and sidelobes, not unlike the patterns observed in the measured vs. model beam plots in Figures \ref{fig:zenithtilemap}, \ref{fig:E03S00tilemap}, \ref{fig:W03S00tilemap}. In particular, a slight main lobe widening is observed in the EW beams and a slight $\sim0.5^\circ$ tilt is observed in the NS beams. These discrepancies are observed across all three pointings, suggesting they are due to some combination of the tile non-idealities discussed above as opposed to per-pointing gain and delay errors in the beamformer.

\section{Discussion}
\label{sec:discussion}

We have used the ORBCOMM satellite constellation to test of the feasibility of a sky transmitter-based beam measurement system for low frequency radio interferometers. Our system compares the power received by an AUT to that from a well-modeled reference dipole, whose ratio gives the relative beampattern of the AUT. The $\sim30$ ORBCOMM satellites provide sky coverage over two thirds of the visible sky from our Green Bank site in a single day, in part due to their low earth orbits and quick orbital precession. Their bright signals probe deep into antenna sidelobes, yielding measurements over 30 dB of MWA tile dynamic range, even after rejecting all data within 20 dB of the galactic noise background. This is an order of magnitude improvement in beam measurement depth over recent source-based beam measurements \citep{colegate14}.

We find through our null experiment that we are limited by 5\% statistical scatter within $20^\circ$ of zenith, and $10\%$ systematics farther out. We hope to definitively identify these fluctuations as multipath scattering in the future using multi-frequency probe signals to investigate their frequency dependence. The time scale of multipath fluctuations is set by the satellite's motion through the frequency-dependent interference pattern set up on the ground by nearby reflecting structures. Finer antenna alignment at the sub-degree level will also help mitigate systematics near the edges of the main lobe. 

We have used this prototype system to conduct the first measurements of an MWA tile beampattern over a large field of view including the main lobe and primary sidelobes. We find good agreement with a simple array factor-based short dipole model within the tile FWHM (23 degrees across at 137 MHz), but observe $\sim$ dB level systematic deviations towards the edge of the main lobe ($\theta\sim20^\circ$) and in the sidelobes. These deviations are larger than the $10\%$ systematics observed in our null experiment, and in principle represent the beam modeling errors we originally sought to measure. 

However, several considerations prevent our interpretation of these deviations as inherent in the MWA tile design, and thus representative of the 128 deployed antennas at the Murchison Radio Observatory in Western Australia. \citet{sutinjo2015} show MWA imaging results which are consistent with an advanced MWA tile model including mutual coupling, however our measured beampatterns appear no more consistent with this model than with our simple analytic one. Indeed the deviations we observe lack the symmetry expected from a beam modeling error due only to insufficiently precise modeling of the sort the cross coupling model is designed to correct. In that case, the zenith pointing EW and NS beampatterns should coincide after a $90^\circ$ rotation, and each should exhibit $180^\circ$ rotational symmetry. Additionally, as discussed in Sec. \ref{sec:mwaerrors}, known MWA beamforming errors are predicted to be at most $\sim20\%$ towards the edge of the main lobe and in sidelobes. However, there exist additional errors peculiar to our MWA tile set up in Green Bank which are more difficult to quantify (e.g., tile and dipole tilts, imperfect ground, ground screen sag, electric potential non-constancy). We thus view this work less as a measurement of \textit{the} MWA tile beampattern and more as a demonstration of the power of the ORBCOMM technique for identifying beam modeling errors (i.e., deviations of an as-built antenna from from its ideal model) down to the -30 dB level of the beam. We expect that further tests using probe signals from drones or multi-frequency satellites will both hone our understanding of the technique and set tighter constraints on beam models of the MWA tile.

Still unknown, though, are the effects of these errors on ongoing MWA Epoch of Reionization power spectrum measurements, which will depend both on their frequency dependence and the degree to which they limit calibration fidelity. This work has begun to probe these effects in a way that imaging cannot, as beamforming errors tend to average out when forming an image with many MWA tiles. However, such averaging of beamforming errors will be less perfect when forming sky power spectra because different antennas probe different regions of the $uv$ plane, depending on which baselines they are part of. Indeed as proposed by \citet{moralesandmatejek}, interferometric imaging algorithms taking per-antenna primary beams into account \citep{fhd, dillonmapmaking} may be critical in order to access the Epoch of Reionization.

\appendix
\section{Measurement of the unpolarized beampattern}
\label{sec:measurementappendix}

In general the response of an antenna to unpolarized radiation (e.g., thermal emission) is different from its response to polarized radiation (e.g., satellite signals). We show how the unpolarized beampattern (defined below) of the AUT may be measured despite any polarization of the satellite probe signal, assuming that both the AUT and the reference dipole have the same polarization response. 

In general, the voltage responses of these antennas to radiation from ($\theta,\phi$) relate to the two incident sky polarizations as
 \begin{eqnarray}
\vec{V}_\mathrm{AUT}&=&\textbf{A}\textbf{R}\vec{E}\label{eqn:measaut} \\
\vec{V}_\mathrm{ref}&=&\textbf{R}\vec{E}\label{eqn:measref}.
\end{eqnarray}
where $\vec{V}=\bigl(\begin{smallmatrix}V_x\\ V_y\end{smallmatrix} \bigr)$ and $\vec{E}=\bigl(\begin{smallmatrix}E_\theta \\ E_\phi\end{smallmatrix} \bigr)$. We use coordinates where $\hat{x}$ points to the East, and $\hat{y}$ points to the North. Spherical unit vectors $\hat{\theta}$ and $\hat{\phi}$ point in the directions of increasing $\theta$ and $\phi$, respectively. These unit vectors are thus functions of those angles, though here we consider radiation incident only from the single direction ($\theta,\phi$). Further, $\textbf{R}$ is a matrix which converts from sky polarization to instrument polarization and drops the $\hat{z}$ component (it is a subset of a rotation matrix),
\begin{equation}
\label{eqn:rotmat}
\textbf{R}=\left(\begin{array}{ccc}
\cos\theta\sin\phi & \cos\phi\\
\cos\theta\cos\phi & -\sin\phi
\end{array}\right).
\end{equation}

Assuming both antennas have the same polarization response, in the sense that the $\hat{x}$ ($\hat{y}$) oriented dipoles respond only to $\hat{x}$ ($\hat{y}$) polarized radiation, $\textbf{A}$ is a diagonal matrix which is a function only of angle on the sky and does not mix instrument polarizations. The physical origin of $\textbf{A}$ is the array factor of the MWA tile, as well as effects of ground screen, antenna geometry, and any dipole cross-coupling. In the ideal case, $\textbf{A}_{11}(\theta,\phi)=\textbf{A}_{22}(\theta,\phi+\pi/2)$, though our measurement does not assume this.

Consider the $\hat{x}$ oriented antennas as an example. The received powers in response to polarized radiation are
\begin{eqnarray}
P_\mathrm{AUT,x}&=& |\textbf{A}_{11}|^2|[R_{11}^2\langle|E_\theta|^2\rangle+2R_{11}R_{12}\Re\langle E_\theta E_\phi*\rangle \nonumber \\
&&+R_{12}^2\langle|E_\phi|^2\rangle]\label{eqn:poweraut} \\
P_\mathrm{ref,x}&=&R_{11}^2\langle|E_\theta|^2\rangle+2R_{11}R_{12}\Re\langle E_\theta E_\phi*\rangle\nonumber \\
&&+R_{12}^2\langle|E_\phi|^2\rangle.\label{eqn:powerref}
\end{eqnarray}

However, if the incident radiation is unpolarized with intensity $I$, then $E_\theta$ and $E_\phi$ are uncorrelated and equal in power and may be added together as powers to give the total received power ($I\equiv2\langle|E_\theta|^2\rangle=2\langle|E_\phi|^2\rangle$). The received powers are then
 \begin{eqnarray}
P_\mathrm{AUT,x}&=& |A_{11}|^2|(R_{11}^2+R_{12}^2)I/2 \\
P_\mathrm{ref,x}&=&(R_{11}^2+R_{12}^2)I/2,
\end{eqnarray}
and so the unpolarized $x$ beampatterns are given by
\begin{eqnarray}
B_\mathrm{AUT,x}&=& |A_{11}|^2|(R_{11}^2+R_{12}^2)/2\label{eqn:autbeam} \\
B_\mathrm{ref,x}&=&(R_{11}^2+R_{12}^2)/2\label{eqn:refbeam}.
\end{eqnarray}

To measure $B_\mathrm{AUT,x}$, we thus need only $A_{11}$, which can be computed from the ratio of $P_\mathrm{AUT,x}$ and $P_\mathrm{ref,x}$, as $\textbf{R}$ is known (see Eqns. \ref{eqn:poweraut} and \ref{eqn:powerref}). This is exactly the procedure outlined in Sec. \ref{sec:processingsatellitepasses}.  

\begin{acknowledgments}
This work was supported by NSF grant AST-0821321, the Marble Astrophysics Fund, and the MIT School of Science. We thank Pat Klima, Bang Dinh Nhan, and the staff at the National Radio Astronomy Observatory--Green Bank for assistance in setting up and debugging our experiment, and Haoxuan Zheng, Aaron Ewall-Wice, Josh Dillon, Lu Feng, and Daniel Jacobs for helpful discussions. We thank Nithyanandan Thyagarajan, Josh Dillon, Adrian Sutinjo, Tim Colegate, and the anonymous referees for very helpful comments on our manuscript.

This scientific work makes use of the Murchison Radio-astronomy Observatory, operated by CSIRO. We acknowledge the Wajarri Yamatji people as the traditional owners of the Observatory site. Support for the MWA comes from the U.S. National Science Foundation (grants AST-0457585, PHY-0835713, CAREER-0847753, and AST-0908884), the Australian Research Council (LIEF grants LE0775621 and LE0882938), the U.S. Air Force Office of Scientific Research (grant FA9550-0510247), and the Centre for All-sky Astrophysics (an Australian Research Council Centre of Excellence funded by grant CE110001020). Support is also provided by the Smithsonian Astrophysical Observatory, the MIT School of Science, the Raman Research Institute, the Australian National University, and the Victoria University of Wellington (via grant MED-E1799 from the New Zealand Ministry of Economic Development and an IBM Shared University Research Grant). The Australian Federal government provides additional support via the Commonwealth Scientific and Industrial Research Organisation (CSIRO), National Collaborative Research Infrastructure Strategy, Education Investment Fund, and the Australia India Strategic Research Fund, and Astronomy Australia Limited, under contract to Curtin University. We acknowledge the iVEC Petabyte Data Store, the Initiative in Innovative Computing and the CUDA Center for Excellence sponsored by NVIDIA at Harvard University, and the International Centre for Radio Astronomy Research (ICRAR), a Joint Venture of Curtin University and The University of Western Australia, funded by the Western Australian State government. Data on which figures and tables herein are based may be obtained by contacting the corresponding author Abraham Neben (abrahamn@mit.edu).
\end{acknowledgments}

%
%
%
%
%
%
%
%
%

\bibliographystyle{agufull08}



%

%
%
\end{article}
%
%
%
%
%

\end{document}